\documentclass[conference]{IEEEtran}
\IEEEoverridecommandlockouts
\usepackage{amsmath}
\usepackage[utf8]{inputenc}
\usepackage[english]{babel}
\usepackage[]{amssymb} 
\usepackage{graphicx}
\usepackage{url}
\usepackage{setspace}
\usepackage{xcolor}
\usepackage{caption}
\usepackage{subcaption}
\usepackage{enumitem}
\usepackage{bm}
\newcommand*{\rom}[1]{\expandafter\@slowromancap\romannumeral #1@}
  
\usepackage{algorithm}
\usepackage{algorithmicx} 
\usepackage[noend]{algpseudocode}

\usepackage{listings}
\usepackage{array}
\newcommand{\PreserveBackslash}[1]{\let\temp=\\#1\let\\=\temp}
\newcolumntype{C}[1]{>{\PreserveBackslash\centering}m{#1}}
\newcolumntype{R}[1]{>{\PreserveBackslash\raggedleft}m{#1}}
\newcolumntype{L}[1]{>{\PreserveBackslash\raggedright}m{#1}}

\usepackage[skip=2pt,belowskip=-2pt,aboveskip=6pt]{caption}
\makeatletter
\let\@authorsaddresses\@empty
\makeatother
\usepackage[english]{babel}
\usepackage{blindtext}

\begin{document}
\title{ASTRA: Low-Overhead Runtime Architecture for STReam Adaptation in Video Analytics}

\graphicspath{ {./images/} }

\author{
\IEEEauthorblockN{Mahshid Ghasemi, Zoran Kostic, Javad Ghaderi, and Gil Zussman}
\IEEEauthorblockA{
Electrical Engineering, Columbia University \\
\{mahshid.ghasemi, zk2172, jghaderi, gil.zussman\}@columbia.edu
}
}
\pagestyle{plain}
\maketitle

\begin{abstract} 
Real-time video analytics is crucial for smart city applications and cloud-connected vehicle control. 
To improve analytics accuracy, it is desirable to process the video at the highest resolution and frame rate. However, due to limited resources, streaming and processing video at the highest resolution and frame rate from all cameras is not feasible and adversely affects the analytics latency. Intelligent adaptation of cameras' resolutions and frame rates based on network conditions and the video content is crucial in order to optimize the performance. In this paper, we present ASTRA, a low-overhead runtime architecture for online adaptation of live camera analytics at the edge. ASTRA can execute various online algorithms as a black box. We deployed ASTRA in the realistic NSF COSMOS testbed and uniquely assessed its real-time performance using COSMOS' street-level cameras. We further evaluated ASTRA with up to eight emulated cameras by streaming a comprehensive video dataset under real-world network conditions. We used ASTRA’s architecture to evaluate the practical performance of several classes of adaptation algorithms, including theoretical and empirical methods. The results indicate that ASTRA can provide system reliability (i.e., the probability of meeting accuracy and latency requirements) of more than 90\% while maintaining performance within a deviation of less than 10\% from optimal offline performance with average GPU utilization overhead of around 2\% per camera.
\end{abstract}
\maketitle 
\thispagestyle{empty}
\section{Introduction}\label{sec:intro}
Video analytics, powered by Deep Neural Networks (DNNs), has gained importance in a wide range of applications, including smart city infrastructure, traffic signal optimization, and cloud-connected vehicle control~\cite{ghasemi2022real,ghasemi2023traffic,ghasemi2025pave}. The recent proliferation of traffic cameras, coupled with the availability of advanced edge/cloud computing resources, is expected to facilitate large-scale video analytics. 

\begin{figure}[t]
\centering
\includegraphics[width=1\columnwidth]{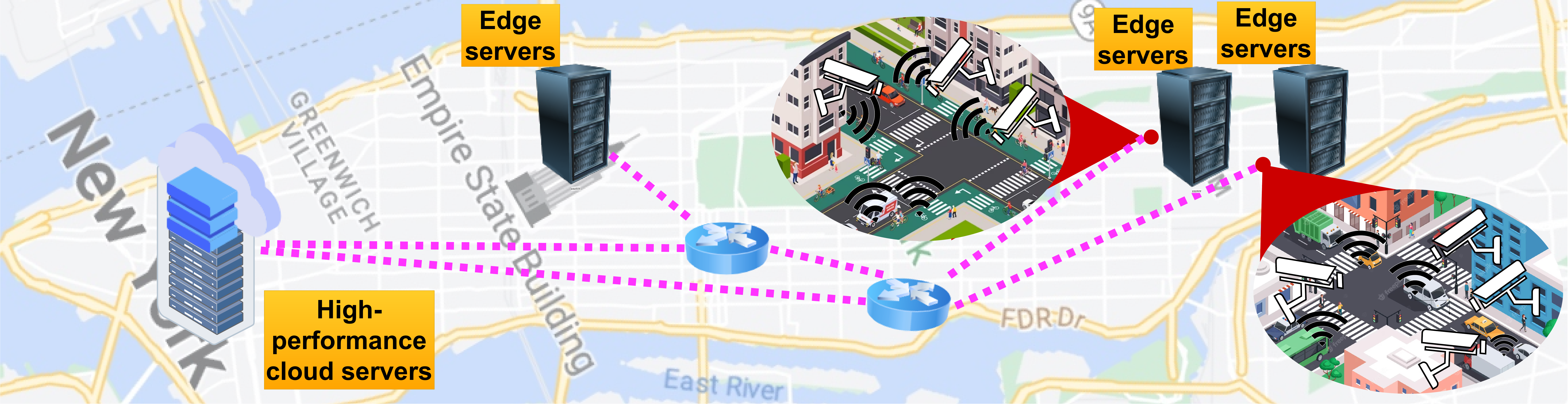}
\caption{The NSF COSMOS testbed uses geo-distributed cameras and edge servers. It allows emulating real-world scenarios that require resource allocation for analyzing real-time video streams~\cite{cosmos-cameras, ghasemi2023traffic,raychaudhuri2020challenge}.}
\label{fig:cosmos pilot site}
\vspace*{-6pt}
\end{figure}

Consider a system of geo-distributed cameras streaming to edge/cloud servers for real-time object detection. These cameras continuously generate a large volume of data that must be encoded, transmitted, decoded, and processed in real-time while adhering to network and computational capacity constraints. A microcosm of such a system is deployed in the NSF COSMOS wireless edge-cloud testbed in New York City (NYC) (see Fig.~\ref{fig:cosmos pilot site})~\cite{raychaudhuri2020challenge,ghasemi2023traffic}. As part of this testbed, several traffic cameras are installed at busy streets and intersections in Manhattan, and their real-time video streams are processed at COSMOS' edge servers to extract traffic- and crowd-related information such as pedestrian and vehicle density. 

Two key performance metrics in such systems are \textbf{accuracy} and \textbf{(end-to-end) latency}. \textbf{\textit{Accuracy}} is determined by evaluating how closely the output of the object detection (i.e., bounding boxes and labels) aligns with manually annotated ground-truth data. \textbf{\textit{Latency}} comprises encoding latency, network latency, decoding latency and DNN inference latency. Accuracy and latency are significantly impacted by network conditions, computational capacity, and video content.
\begin{figure*}[t]
\centering
\includegraphics[width=0.9\linewidth]{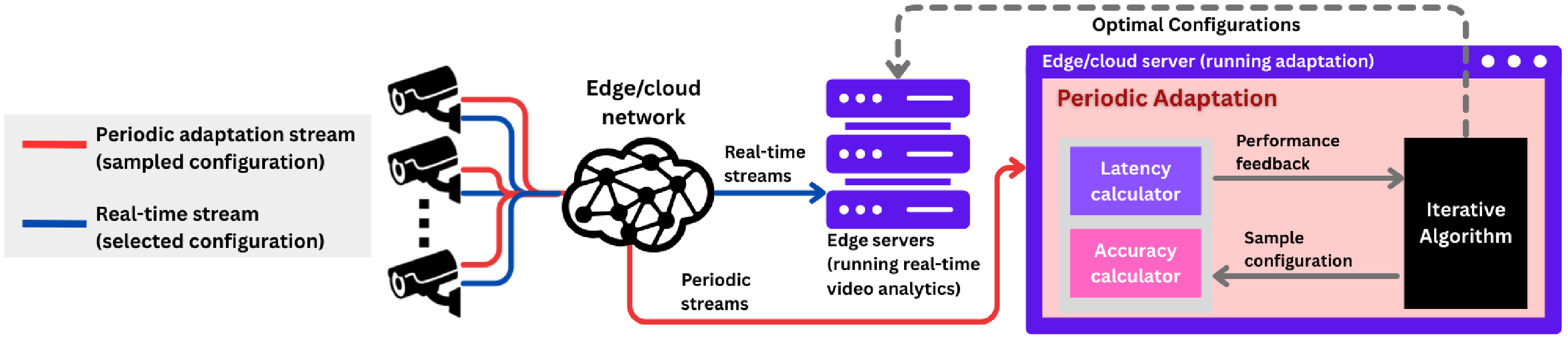}
\caption{Deployment of online adaptation in a multi-camera setup connected to edge servers. The adaptation system updates the configuration of the system periodically in response to changes in the environment.}
\vspace*{-10pt}
\label{fig:astra-pipeline-hl}
\end{figure*}

\noindent\textbf{Network conditions.} Network variations can cause lower throughput, increased packet loss, and buffering. Therefore, if in such cases, resolution or frame rate are not reduced, (i) network latency increases, and/or (ii) quality of decoded images degrades which reduces accuracy.

\noindent\textbf{Computational capacity.} Real-time processing of video streams using DNNs is computationally demanding. For example, the execution of a conventional object detection model on a single 1K resolution video stream at 30\thinspace{fps} can consume most of the processing capacity of an edge GPU. In such a case, processing additional streams leads to increased inference latency unless the resolution or frame rate of at least one of the video streams or the DNN model size is reduced. 

\noindent\textbf{Video content.} Depending on the density level and movements in the monitored scene, different resolutions, frame rates, and DNN complexities are required in order to achieve sufficient accuracy. For example, for cameras viewing urban streets, when vehicles are moving slowly (e.g., at a red traffic light or during heavy traffic) the frame rate can be reduced without losing accuracy. Similarly, when the pedestrians' density is low, or they are close to the camera, the resolution can be reduced without affecting the accuracy.

Therefore, optimizing parameters such as resolution and frame rate when feasible, can significantly save on bandwidth and computational power. 

\vspace*{1pt}
\noindent\textbf{Iterative adaptation algorithm.} In this paper, the objective is to dynamically identify video streams' resolutions and frame rates that maximize the overall performance (a function of accuracy and end-to-end latency), while meeting application-specific constraints on minimum accuracy and maximum end-to-end latency. We refer to a specific combination of resolution and frame rate for a stream as a \textbf{\emph{configuration}}.  

Our measurements, consistent with prior research~\cite{zhang2022casva, khani2023recl, jiang2018chameleon, galanopoulos2021automl}, indicate that accuracy and latency are \textit{time-varying and unknown} functions of the configuration. The characteristics of these functions depend on dynamic variables such as video content, available network and computing resources, and DNN models. Identifying an optimal configuration through deterministic or offline optimization with empirical modeling \textit{cannot} capture the unknown and time-varying nature of the problem and often leads to poor performance. This necessitates iterative online adaptation in such unpredictable settings, where configurations can be evaluated only when deployed and processing real video streams, without knowledge of the future or an explicit model for the accuracy or latency functions. At each iteration, such algorithms select a configuration to evaluate based on previously observed performance. We refer to such algorithms as \textit{\textbf{iterative algorithms}}.

\vspace*{1pt}
\noindent\textbf{Key idea.} The primary goal of online adaptation is to maximize performance under constrained resources. Achieving this requires that: (i) the adaptation runtime architecture incurs minimal overhead when executing the iterative algorithm, and (ii) given the rapid evolution of online adaptation algorithms, it is crucial that the architecture allows for easy integration of new algorithms, requiring only minimal changes. To meet these requirements, we introduce \textbf{ASTRA}, a low-overhead runtime \underline{\textbf{A}}rchitecture for \underline{\textbf{STR}}eam \underline{\textbf{A}}daptation in real-time video analytics that enables execution of existing and future iterative algorithms as a black-box (as shown in Fig.~\ref{fig:astra-pipeline-hl}). A patent based on ASTRA is pending~\cite{ghasemi2026patentastra}.

\vspace*{1pt}
\noindent\textbf{Prior studies.} 
Significant work has been done on exploring online video analytics adaptation approaches. However, most previous studies focus solely on developing adaptation algorithms (the “content” of the black-box in Fig.~\ref{fig:astra-pipeline-hl}) and are evaluated using recorded videos, not within an end-to-end real-time system~\cite{jiang2018chameleon,zhang2022casva,nigade2022jellyfish, bhardwaj2022ekya,faye2024videojam}. Despite the growing number of online adaptation algorithms for video analytics, there remains a significant gap in understanding the \emph{system-level overheads} and \emph{practical feasibility} of deploying such algorithms within an edge/cloud network and operating them in real-time. ASTRA is designed to close this gap. To our knowledge, ASTRA is the first to provide a detailed architectural design and practical implementation for executing iterative adaptation algorithms in a real-time video analytics pipeline.

\subsection{Challenges}\label{sec:challenges}
\noindent\textbf{Architecture challenges.} Two main obstacles in executing online adaptation algorithms are (i) the cost of performance measurements at each iteration and (ii) the delay incurred in iterative switching between configurations and video streams. 

\textbf{(i)\textit{ Performance measurement cost}}: Online adaptation algorithms require feedback that indicates the observed performance of a specific configuration. For example, for object detection, to measure the accuracy associated with a given configuration, the detected objects (i.e., bounding boxes and their labels) obtained with that configuration must be compared against the ground truth. Since it is impossible to access manually annotated ground truth in real-time, a \textit{\textbf{proxy ground truth}} obtained via the most resource-intensive configuration (highest resolution and frame rate) may be used instead. Generating proxies, by definition, requires significant computational and network overhead, and so does measuring the accuracy/latency of a given configuration.

\vspace*{2pt}
\textbf{(ii)\textit{ Switching delay}}: Executing iterative adaptation requires evaluating various configurations by sequentially running object detection models on short segments of the video stream under each configuration. This necessitates switching between video streams and analytics pipelines with different configurations. Each switching introduces non-trivial overhead that can significantly increase adaptation time. This overhead includes repeated decoding and encoding, buffering, pipeline reconfiguration, data transfer (e.g., from storage to memory or CPU to GPU), cache inefficiencies, and model loading. In real-time systems, even minor overheads, when recurrent, accumulate and compromise performance. Therefore, to ensure practical feasibility and scalability, these overheads must be mitigated.

\noindent\textbf{Algorithmic challenges.} 
Due to the temporal and computational overhead of each iteration of online adaptation, the iterative algorithm must be capable of attaining a near-optimal configuration in a minimum number of iterations to ensure near-optimal performance most of the time.

\subsection{Contributions} 
\noindent\textbf{Identifying and measuring runtime overheads in online video adaptation.} We performed extensive profiling across memory, CPU, and GPU resources, and identified hidden bottlenecks and overheads in online adaptation systems. Quantifying these overheads and bottlenecks is crucial for verifying the feasibility of deployment. 

\noindent\textbf{Efficient architecture design to mitigate runtime overheads.} We designed and implemented ASTRA, an online adaptation framework that mitigates the identified overheads. Starting from a baseline design, we systematically refined the architecture to reduce inefficiencies. ASTRA incorporates asynchronous processing, OS-level signaling, efficient memory management, and modular design. ASTRA enables performance evaluation and fair comparison of different iterative algorithms within a unified framework.


\noindent\textbf{Executing representative adaptation algorithms as interchangeable black-box components.} To compare different algorithmic choices, we used ASTRA to execute two iterative algorithms (the ``content" of the black-box in Fig.~\ref{fig:astra-pipeline-hl}): (i) the Gaussian Process Upper Confidence Bound with Constraints (GP-UCB-C) algorithm derived from~\cite{xu2023constrained}, and (ii) Chameleon++ derived from~\cite{jiang2018chameleon}. These algorithms were chosen to represent two categories of common iterative adaptation algorithms: (1) theoretically guaranteed algorithms (GP-UCB-C), and (2) empirical approximated algorithms (Chameleon++). We then compared their performance against two baselines: (a) the offline optimal (upper bound on achievable performance) and (b) a ground-truth-free adaptation method (AutoML++) that estimates accuracy via confidence scores instead of ground-truth.
\begin{figure}[t]
\centering
\includegraphics[width=1\linewidth]{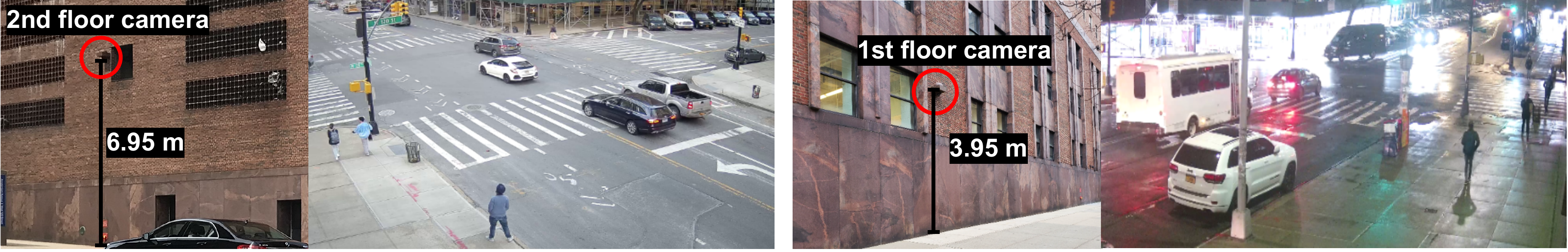}
\caption{COSMOS cameras deployed on the 1\textsuperscript{st} and 2\textsuperscript{nd} floor of a building and their view of 120$^\textrm{th}$ St.\ and Amsterdam Ave.\ intersection, NYC~\cite{cosmos-cameras, yang2020cosmos}.}
\label{fig:cameras-view}
\vspace*{-10pt}
\end{figure}

\noindent\textbf{Realistic evaluation:} We evaluated ASTRA in two complementary settings designed to capture both real-world conditions and fully controlled, reproducible scenarios.

\noindent\textbf{\emph{(i) Deployment on the COSMOS testbed.}} We deployed ASTRA in the COSMOS testbed network and performed \emph{real-time} adaptation for two live traffic cameras (depicted in Fig.~\ref{fig:cameras-view}) deployed on the 1$\textsuperscript{st}$ and 2$\textsuperscript{nd}$ floor of a building viewing an intersection in NYC~\cite{cosmos-cameras}.

\noindent\textbf{\emph{(ii) Controlled and reproducible evaluation.}} A fair comparison among algorithms necessitates using identical and reproducible conditions. To achieve this, we deployed ASTRA on Google Cloud virtual machines (VMs) and conducted experiments with eight emulated cameras under realistic network latency conditions. Cameras were emulated by streaming videos via the Real-time Transport Protocol (RTP)~\cite{schulzrinne2003rfc3550} using GStreamer~\cite{taymans2013gstreamer}. Unlike prior studies (e.g.,~\cite{du2023oneadapt,wong2024madeye,zhang2024starstream,wang2023shoggoth, zhao2025mocha}) that rely solely on pre-recorded videos, our emulated setup preserves the essential characteristics of live deployments while also enabling reproducibility and identical conditions across algorithms.

\noindent\textbf{\bfseries Ethical consideration.} The use of COSMOS' live and recorded video streams~\cite{cosmos-cameras,yang2020cosmos} was designated IRB-exempt by Columbia University. These videos are utilized solely for academic and research purposes and will be distributed only after appropriate anonymization, including obscuring faces and license plates. More generally, the authors had several discussions with local community stakeholders, which point to the fact that the use of videos for improved traffic flow and safety (a common use by municipalities) is acceptable by the community.

\noindent\textbf{Results.} Compared to baseline architectures, ASTRA reduces the per-iteration execution overhead of the iterative adaptation algorithm by an order of magnitude (e.g., from seconds to hundreds of milliseconds). ASTRA, when integrated with the GP-UCB-C algorithm, achieves performance within \textbf{10\%} of the optimal performance. It ensures system reliability (i.e., probability of meeting the accuracy and latency requirements) of more than \textbf{90\%}. ASTRA incurs low adaptation overhead, averaging around \textbf{2\%} GPU utilization per camera over time.

\section{Related Work}\label{sec:related-work}
Several studies presented various methods to enhance the performance of real-time video analytics despite the dynamics in the environment, such as network conditions variation, video content (e.g., density, lighting, and weather conditions) variations. Related work primarily falls into the following categories based on their optimization approaches.

\noindent \textbf{Configuration adaptation.} Studies employ periodic adaptation, including DNN retraining or continuous updates~\cite{bhardwaj2022ekya,kong2023edge, khani2023recl,wang2023shoggoth,zhao2025mocha}, dynamic model switching~\cite{du2023oneadapt,nigade2022jellyfish}, and adjusting resolution, frame rate, or bitrate~\cite{jiang2018chameleon,wu2021soudain,zhang2018awstream,kim2020lightweight,wang2020joint,galanopoulos2021automl,yi2025arma}. Specialized systems address specific constraints such as packet loss~\cite{yang2024logan}, satellite links~\cite{zhang2024starstream}, or camera orientation~\cite{wong2024madeye}. RIVA~\cite{li2025riva} uses hierarchical reinforcement learning to adapt bitrate allocation and online retraining decisions for real-time industrial video analytics under dynamic network and video conditions. OAVS~\cite{li2024oavs} uses hierarchical reinforcement learning to update bandwidth prediction and bitrate allocation under dynamic network and scene conditions for drone-sourced video analytics. A benchmark provided by~\cite{bala2024ooda} evaluates how decoder and model-size choices affect latency in cloudlet-offloaded drone video analytics.

\noindent \textbf{Distributed inference/learning.} Studies leverage distributed computing via DNN partitioning~\cite{chinchali2018neural,zhao2018deepthings} and efficient edge-cloud resource allocation or load balancing~\cite{bartolomeo2023oakestra,jano2023nextgsim,fu2022split,gao2021edgesp,yang2023javp,liu2022sniper,murad2022dao, wei2023nn, hou2025viedge, faye2024videojam}. Concurrent DNN execution on heterogeneous processors is also explored~\cite{jia2022codl,wei2023nn,ling2022blastnet}. VideoJam~\cite{faye2024videojam} designs a pipeline-level load-balancing architecture for live video analytics. ViEdge~\cite{hou2025viedge} distributes inference across edge resources. The study~\cite{dong2025does} benchmarks on-device versus edge-offloaded real-time vision tasks and shows that edge offload remains important for running high-accuracy models within tight latency constraints. PIB~\cite{fang2024pib} extracts and compresses task-relevant feature maps on edge cameras and fuses them at the edge server. In another study~\cite{sun2024ecoffload}, smart cameras offload tasks to an Unmanned Aerial Vehicle (UAV)-enabled edge server.
    
\noindent \textbf{Video frames filtering.} Studies focus on on-camera frame filtering~\cite{bastani2020miris,moll2022exsample,li2020reducto,paul2021aqua}, region of interest determination~\cite{wang2022vabus,xarchakos2019svq,apicharttrisorn2019frugal,liu2019edge,guo2021crossroi,liu2022adamask,zhang2021elf}, and adaptive encoding~\cite{du2020server,du2022accmpeg,kim2023entro}. Additionally, frame or region-based enhancement techniques are used to improve throughput and accuracy~\cite{lu2022turbo,wang2025regenhance}. RegenHance~\cite{wang2025regenhance} enhances important regions of a frame, achieving higher throughput than naive per-frame super-resolution. JIGSAW~\cite{gokarn2024jigsaw} exploits cross-camera overlap to pack high-utility tiles for efficient edge inference.

Unlike previous studies that primarily focus on developing adaptation methods to maintain video analytics performance under environmental variability, our work centers on the broader \textit{system challenges} of executing configuration adaptation in real-time within a deployed video analytics pipeline. While prior work often models or optimizes specific costs such as bandwidth, latency, or retraining overhead, many practical execution overheads remain underexplored, including switching delays, pipeline reconfiguration, buffering behavior, resource contention, and model switching. Through extensive profiling and measurements, we show that these overheads can significantly affect adaptation performance and, if left unaddressed, can make online adaptation counterproductive by adding latency and computation that degrade overall performance. ASTRA directly targets these challenges: its architecture is explicitly designed to mitigate these real-time execution overheads, enabling online adaptation to deliver meaningful performance gains in real deployments.

\section{Online Configuration Adaptation}\label{sec:online-adaptation}
Consider a video analytics system where real-time streams from several cameras are transmitted to edge servers for object detection using DNN models (see Fig.~\ref{fig:astra-pipeline-hl}). The key performance metrics of such a system are accuracy, (end-to-end) latency, and bandwidth consumption. Bandwidth manifests itself in accuracy and network latency from the camera to the edge server when streaming over Transmission Control Protocol (TCP) connections. As mentioned in Section~\ref{sec:intro}, accuracy and latency are \textit{time-varying, unknown} functions of the system's configurations (e.g., resolution and frame rate of the cameras). The characteristics of these functions depend on dynamic parameters such as video content, model complexity, network conditions, and available computational resources.

\vspace*{2pt}
\noindent\textbf{Adaptation objective.} The objective is to dynamically find the configurations that maximize a value function of accuracy and latency, subject to constraints on maximum latency and minimum accuracy. The unpredictable and dynamic nature of the problem limits the effectiveness of empirical modeling and deterministic optimization in identifying near-optimal configurations. Hence, an iterative algorithm is required to estimate the system behavior in an online manner. At each iteration $t$ of such an algorithm, one \emph{configuration} $x_{t}$ is sampled. Here, $x_{t}\in\mathcal{X}$, where $\mathcal{X}$ is a discrete set consisting of a finite number of supported resolutions and frame rates. Then, $x_{t}$'s achieved \emph{accuracy}, $A_t(x_{t})$, and \emph{latency}, $L_t(x_{t})$ on a short segment (e.g., $\tau$=0.2\thinspace{seconds}) of the camera's video stream is evaluated. The results up to iteration $t$ are used to decide on the subsequent configuration $x_{t+1}$ to be evaluated. The algorithm must be terminated upon identification of a near-optimal configuration. We refer to such algorithms as \emph{\textbf{iterative algorithms}}.

\vspace{2pt}
\noindent\textbf{Performance metrics.} \textit{Accuracy} $A_t(x_t)$ is calculated by comparing the obtained object detection results (i.e., bounding boxes and labels) using configuration $x_t$ on a video segment against the ground truth. Following previous work~\cite{jiang2018chameleon, du2020server,wang2020joint}, we use the F1-score metric. The detection F1-score is computed by verifying if a bounding box shares the same label and has sufficient spatial overlap with the associated ground truth. To account for the impact of frame rate on the accuracy, following the prior research~\cite{jiang2018chameleon,kang2017noscope,zhang2017live}, we use the location of objects from the previous sampled frame for a frame that is not sampled by the configuration. \textit{Latency}, $L_t(x_t)$ is defined as the sum of the encoding latency by the camera, the network latency to transfer video segment from the camera to the server, the decoding latency on the edge server, and the inference latency to run object detection on the video segment with configuration $x_t$. 

\vspace{2pt}
\noindent\textbf{Proxy ground-truth.} The precise ground truth required for calculating accuracy $A_t(x_t)$ can only be obtained through manual annotation, which is infeasible in real-time. Replacing manual annotation with analytical approximations~\cite{kim2020lightweight} or with confidence scores generated by the DNN~\cite{galanopoulos2021automl} introduces \emph{unknown errors} into the accuracy measurements. Our experiments (Section~\ref{sec:evaluation}) show that such imprecise accuracy estimates can lead the system to pick configurations that deviate significantly from the actual optimum. DNN models generally achieve their highest accuracy at the highest resolution and frame rate, which is also the most resource-intensive configuration. We refer to this as the \textbf{\emph{golden configuration}}. Because of this, many prior works use the model's output under the golden configuration as the reference for assessing accuracy under less resource-intensive settings~\cite{zhang2024starstream, zhang2022casva,wu2023ilcas}. Throughout the paper, we refer to the object detector's output under the golden configuration as the \textbf{\emph{proxy ground truth}}, or simply the \textbf{\emph{proxy}}. 
\section{Architecture Design}\label{sec:astra-architecture}
\begin{figure}[t]
\centering
\includegraphics[width=1\linewidth]{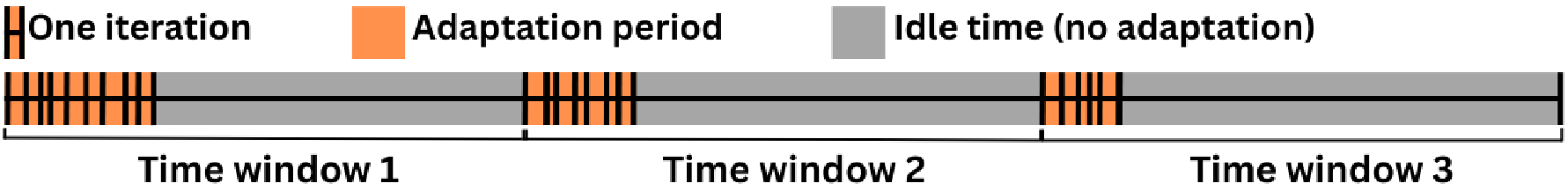}
\caption{Adaptation timeline: The iterative algorithm starts adaptation at the beginning of each time window until a near-optimal configuration is determined, then that configuration is used for the rest of the window.}
\vspace{-15pt}
\label{fig:ideal-adapt-timeline}
\end{figure}

As discussed in Section~\ref{sec:challenges}, each iteration of an iterative adaptation algorithm incurs substantial overhead. In particular, generating proxies (by definition) and sequentially switching between configurations and video streams incur network, computational, and temporal overhead. To execute such iterative algorithms efficiently, these overheads must be minimized. Therefore, an effective adaptation architecture should ensure that the system uses a near-optimal configuration majority of the time while incurring low overhead. Due to \emph{unpredictable} variations in network conditions and video content, the optimal configuration shifts over time, requiring the adaptation process to be invoked \emph{periodically}. The appropriate adaptation frequency depends on environment volatility. We refer to each adaptation cycle as a \textbf{\emph{time window}}. A typical value for the time window is \textbf{1-2 minutes}. As illustrated in Fig.~\ref{fig:ideal-adapt-timeline}, at the beginning of each time window, adaptation resumes until a near-optimal configuration is found by the adaptation algorithm, which is then used for the remainder of the window. To keep adaptation overhead low, the adaptation period should remain small relative to the window length (e.g., below 10\%). In this section, we illustrate how we designed ASTRA, accordingly starting from a baseline design. 
\subsection{Baseline Architectures}\label{subsec:baseline-architec}
\begin{figure*}[t]
\centering
\includegraphics[width=0.8\linewidth]{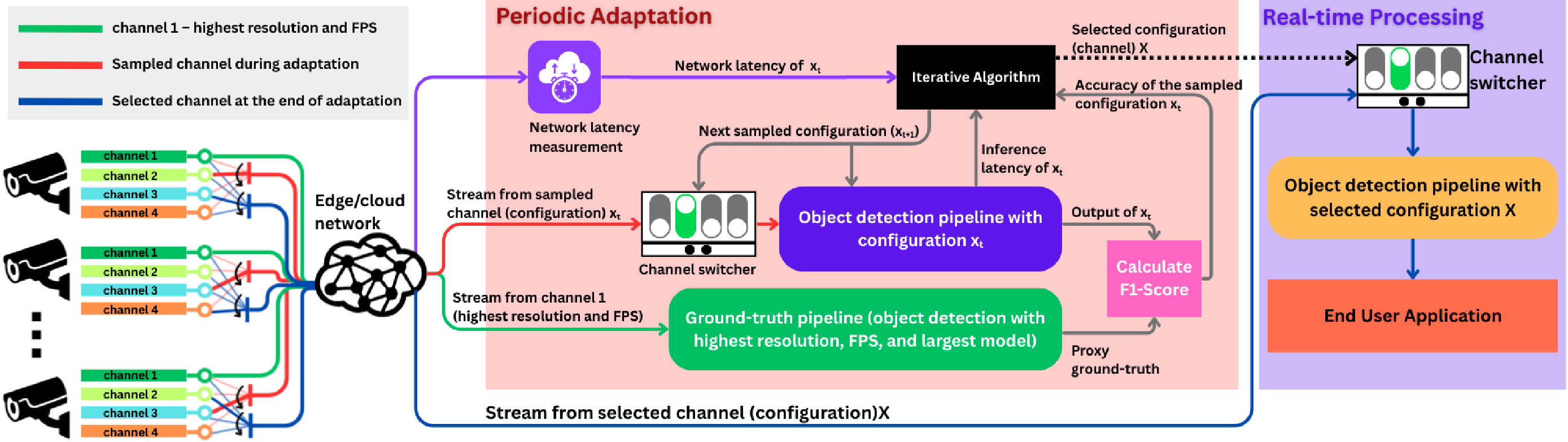}
\caption*{(a) Channel switching: Baseline architecture for online adaptation by executing an iterative algorithm.}
\vspace{1em}
\includegraphics[width=0.8\linewidth]{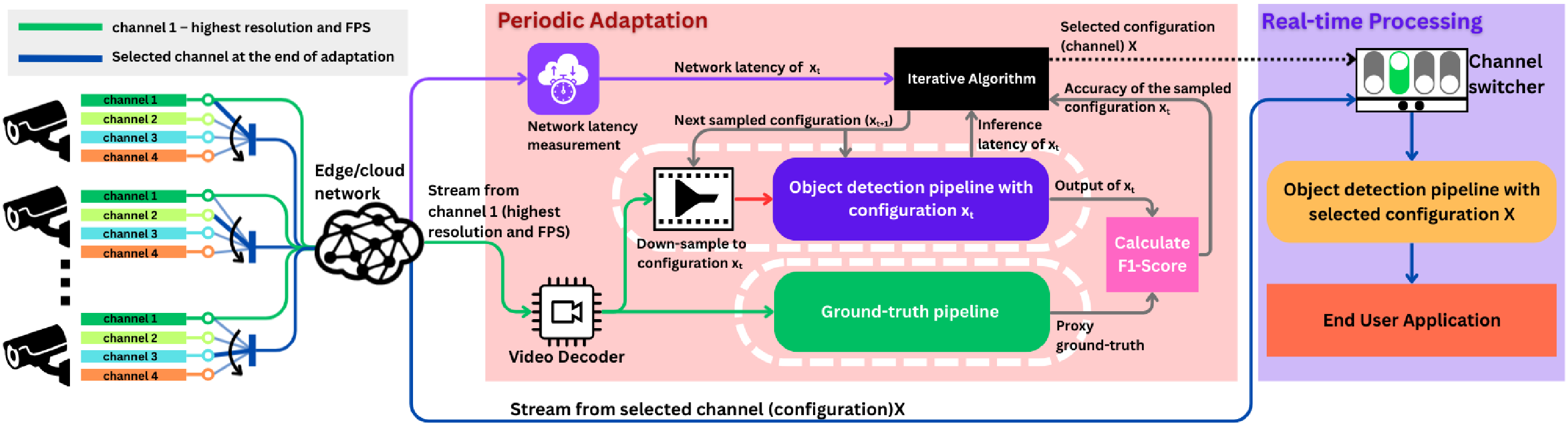}
\caption*{(b) Configuration switching: Improved adaptation architecture that eliminates channel switching delay.}
\caption{Comparison of (a) baseline and (b) improved architectures for online adaptation.}
\label{fig:architecture-comparison}
\end{figure*}

To perform online adaptation by running an \emph{iterative algorithm} described in Section~\ref{sec:online-adaptation}, the baseline architecture for real-time configuration adaptation is shown in Fig.~\ref{fig:architecture-comparison}(a). In this architecture, during an adaptation period for each iteration $t$, the algorithm needs to observe both the accuracy and latency obtained under the \textbf{sampled configuration $x_t$}. This requires the system to momentarily process a short segment of video (for example, a fraction of a second such as 0.2\thinspace{s}) using the channel associated with configuration $x_t$. Therefore, the system needs to switch rapidly between different channels throughout the adaptation period until the algorithm converges to a near-optimal configuration. To understand the feasibility of such rapid switching, we measured channel-switching latency for three types of cameras deployed in the COSMOS testbed: \textbf{HikVision} (DS-2CD5585G0-IZHS 8\thinspace{MP} Outdoor Dome) camera, \textbf{Bosch} (FLEXIDOME 8100i NDE-8704-R 4K UHD Outdoor Network PTRZ Dome) camera, and \textbf{Axis} (Q3628-VE 8\thinspace{MP} Outdoor Dome) camera. This latency is defined as the time from issuing a request for a new channel to the moment the first packet of the first frame from that channel arrives at the edge server. The edge server connects to the cameras via fiber optics, resulting in negligible network delay (under 10\thinspace{ms}), which means this latency primarily reflects the internal switching latency of the cameras. Each camera is programmed with three channels, with the following resolution and frame per second (fps): ((3840$\times$2160), 30\thinspace{fps}), ((1920$\times$1440), 15\thinspace{fps}), ((800$\times$600), 10\thinspace{fps}). 

\begin{figure*}[t]
\centering
\includegraphics[width=0.9\linewidth]{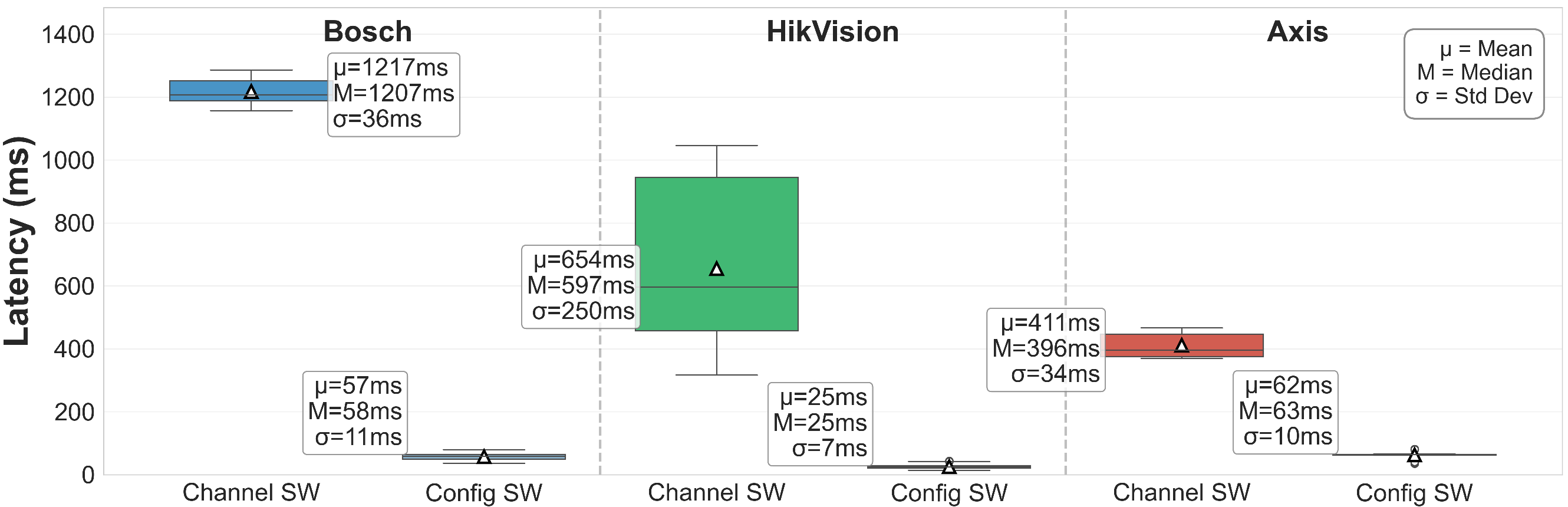}
\caption{Latency comparison of channel switching: transitioning between different camera streams (denoted by Channel SW), and configuration switching: changing resolution/frame rate parameters (denoted by Config SW), corresponding to the architectures in Fig.~\ref{fig:architecture-comparison}(a) and Fig.~\ref{fig:architecture-comparison}(b), respectively.}
\vspace{-10pt}
\label{fig:switching-delay-comp}
\end{figure*}

Fig.~\ref{fig:switching-delay-comp} shows the measured switching latencies for these cameras. The \emph{\textbf{channel switching}} measurements (denoted by Channel SW) are based on around \textbf{100} channel switches per camera, switching among the three available channels (e.g., 1→2, 2→3, 3→1). As Fig.~\ref{fig:switching-delay-comp} indicates, the channel switching latency strongly depends on the camera type. Those cameras (e.g., the Bosch camera) that employ advanced mechanisms and/or larger internal buffers to maintain high-quality and smooth real-time streaming exhibit higher switching delays. These results show that channel switching latency can range from \textbf{$\sim$400\thinspace{ms}} to more than \textbf{1.2\thinspace{s}}, which is significant when the adaptation algorithm needs to switch channels every fraction of a second, such as 0.2\thinspace{s}.

To mitigate this latency and reduce the dependency on camera-specific buffering behavior, we improve upon the baseline architecture shown in Fig.~\ref{fig:architecture-comparison}(a) by introducing the modified design shown in Fig.~\ref{fig:architecture-comparison}(b). We replace the ``\emph{channel switcher}" component that repeatedly requests different channels from the cameras with a \textbf{down-sampler} component. This component only receives the highest-resolution, highest-frame-rate stream (Channel 1, indicated by green arrows in Fig.~\ref{fig:architecture-comparison}) and locally generates other configurations sampled by the iterative adaptation algorithm. Since Channel 1 with golden configuration, hereafter referred to as the \textbf{golden stream}, is already necessary to compute the proxy ground-truth, it can be mirrored internally at the edge server to perform the down-sampling, eliminating the need to request two separate streams from the camera. 

 \vspace*{2pt}
\noindent\textbf{Switching latency.} The effectiveness of the architecture in Fig.~\ref{fig:architecture-comparison}(b) is reflected in the latency measurements shown in Fig.~\ref{fig:switching-delay-comp}. The figure compares two switching approaches across three camera types (Bosch, HikVision, and Axis): \emph{channel switching} (left boxes in each camera section) and \emph{configuration switching} (right boxes in each camera section) achieved through down-sampling. Similarly to the channel switching measurements, the configuration switching measurements are also based on around \textbf{100} down-sampling (configuration switches) from the golden configuration to the other two configurations (i.e., 1$\rightarrow$2 and 1$\rightarrow$3). Fig.~\ref{fig:switching-delay-comp} indicates that the improved architecture in Fig.~\ref{fig:architecture-comparison}(b) could reduce the switching delay by \textbf{approximately 20$\times$}. Importantly, it also mitigates the influence of camera-specific buffering and jitter-control mechanisms, making the system's behavior more predictable across different camera models. 

\noindent\textbf{Bandwidth consumption.} The design in Fig.~\ref{fig:architecture-comparison}(b) also reduces network bandwidth usage by removing the additional camera streams required by the architecture in Fig.~\ref{fig:architecture-comparison}(a), denoted by red arrows.
\begin{figure*}[t]
\centering
\begin{minipage}[t]{0.41\linewidth}
    \vspace{0pt}
    \includegraphics[width=\linewidth]{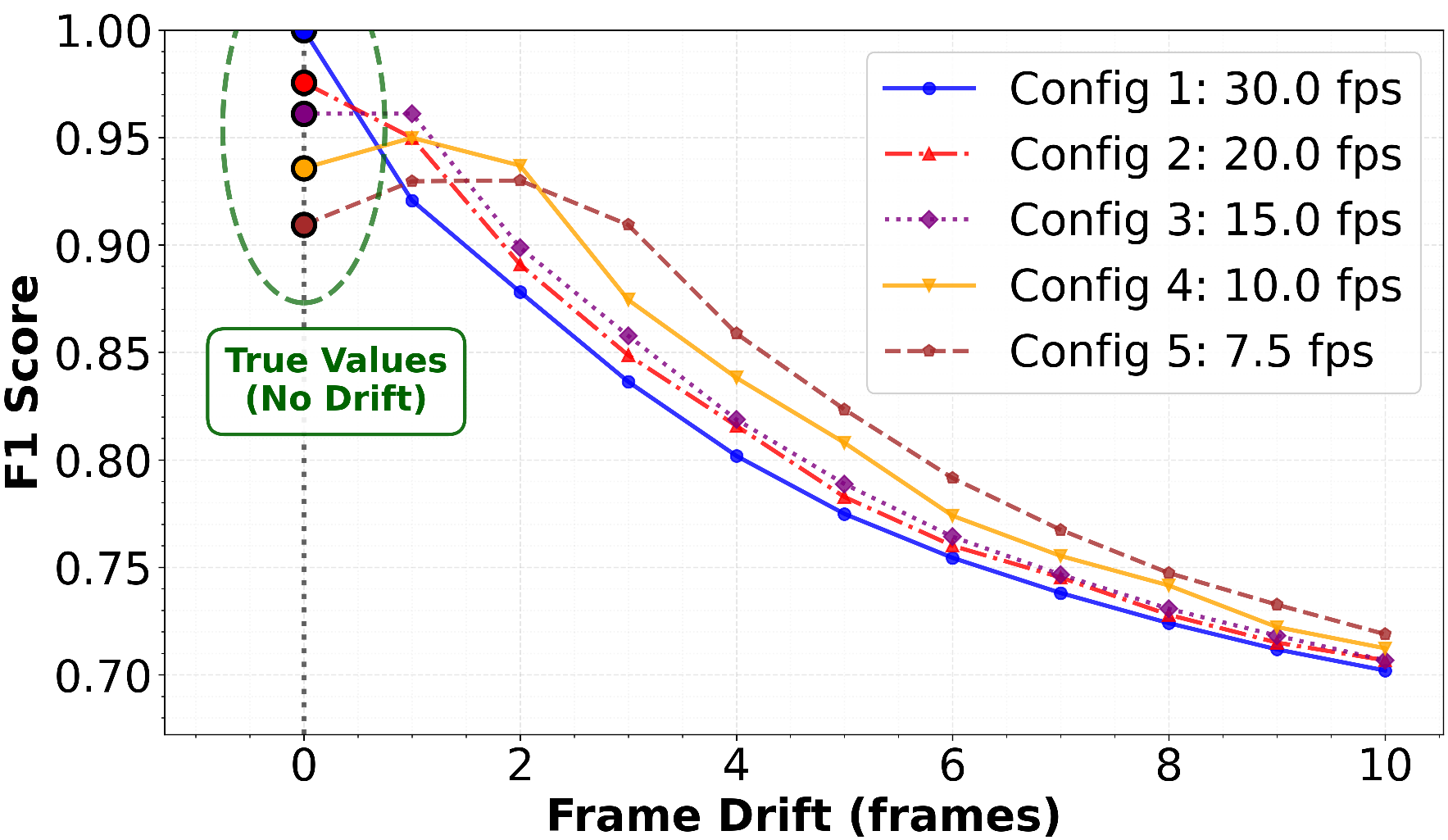}
    \vspace{0.3em}
    \centering
    {\small (a) Resolution: 3840$\times$2160}
\end{minipage}
\hspace{0.1 in}
\begin{minipage}[t]{0.41\linewidth}
    \vspace{0pt}
    \includegraphics[width=\linewidth]{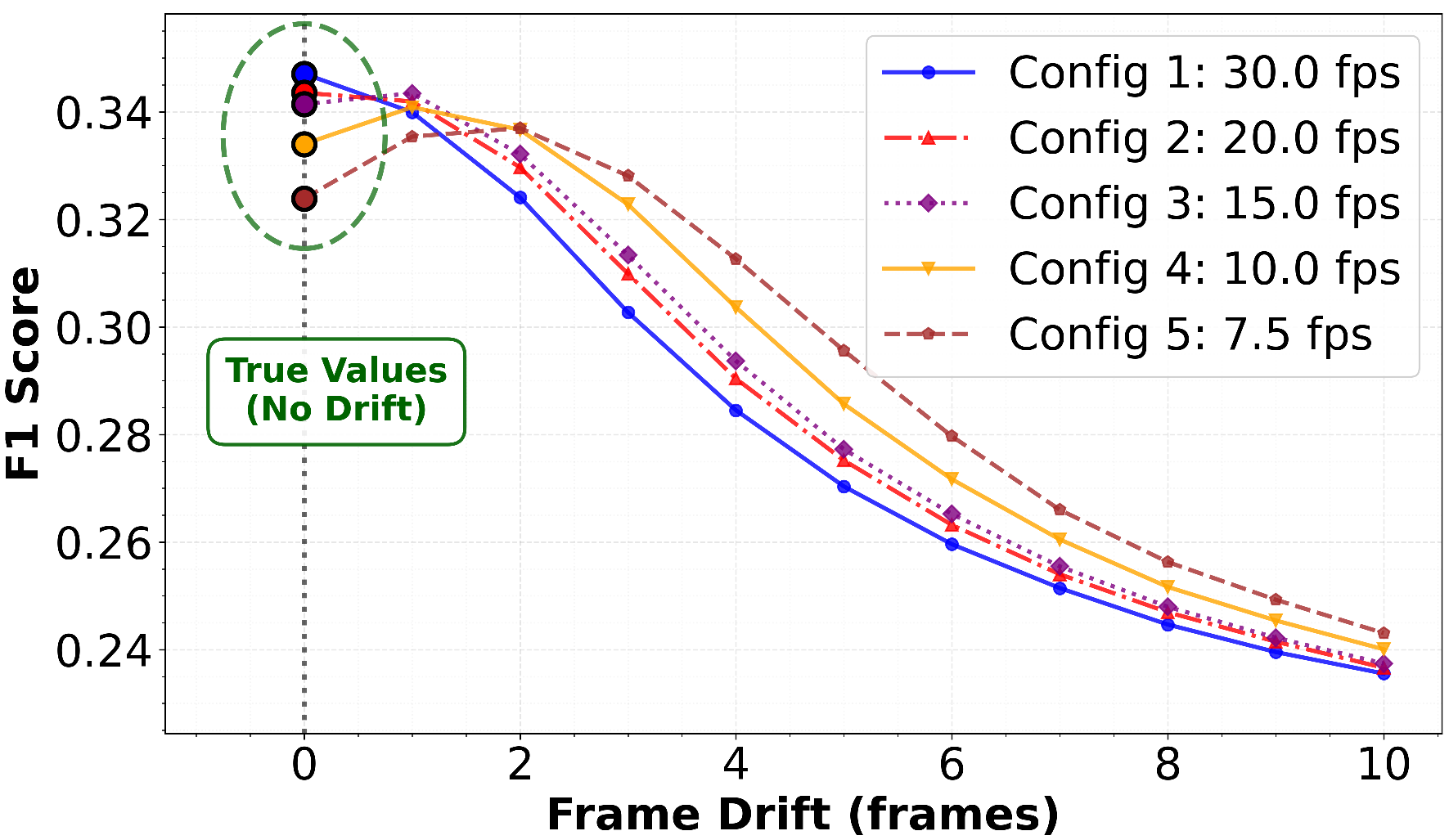}
    \vspace{0.3em}
    \centering
    {\small (b) Resolution: 960$\times$540}
\end{minipage}
\caption{Impact of frame-drift between the \emph{ground-truth pipeline} and the \emph{object detection pipeline} on F1-score measurement for two resolution levels: (a) 3840$\times$2160, and (b) 960$\times$540, with different frame rates.}
\label{fig:frame-drift-impact}
\vspace*{-10pt}
\end{figure*}

\noindent\textbf{Precise F1-score calculation.} The improved architecture in Fig.~\ref{fig:architecture-comparison}(b) also enhances synchronization between the \emph{ground-truth pipeline} (green box) and the \emph{object detection pipeline} (dark purple box). By using a single stream from Channel 1 instead of two separate camera streams, both pipelines operate on identically timestamped frames. This eliminates the need to align frames originating from different channels, which often have inconsistent timing due to independent buffering and Network Time Protocol (NTP) drifts. In some cameras, Real-Time Control Protocol (RTCP) packets, which normally carry timing and synchronization information for media streams, are either absent or incomplete. Consequently, the timestamps embedded in frame packets are not sufficiently reliable for exact frame matching between the two pipelines, making it difficult to compute frame-level accuracy metrics such as the F1-score. The single-stream design solves this issue, ensuring accurate temporal alignment and reliable accuracy calculation.

\noindent\textbf{Importance of temporal alignment.} To quantify the impact of temporal misalignment between the \textit{ground-truth pipeline} and the \textit{object detection pipeline} on accuracy calculation during adaptation, we analyzed a \textbf{10-minute} recording from one of the COSMOS cameras (2\textsuperscript{nd}-floor view in Fig.~\ref{fig:cameras-view}). For each configuration, we computed the average calculated F1-score over consecutive adaptation iterations of \textbf{0.2\thinspace{s}} video duration. The results are shown in Fig.~\ref{fig:frame-drift-impact}.

Accurate F1-score measurement is critical for the \textit{adaptation algorithm} to correctly model the relationship between configuration and accuracy. Even small frame misalignments between the two pipelines can distort this relationship. As shown in Fig.~\ref{fig:frame-drift-impact}, a drift of just one frame can substantially alter the computed F1-score. For example, in both resolution settings (Figs.~\ref{fig:frame-drift-impact}(a) and (b)), although the highest frame rate configuration truly achieves the best accuracy, a minor frame drift causes it to appear as the lowest-performing configuration. Such distortion can mislead the iterative algorithm, resulting in the selection of a suboptimal configuration and defeating the purpose of online adaptation.
\subsection{Our Architecture: ASTRA}\label{subsec:astra-architect}
The architecture should also manage differences in computational and time requirements among concurrent processes effectively to reduce their waiting (idle) time and ensure efficient resource utilization. For example, in Fig.~\ref{fig:architecture-comparison}(b), two major processes (outlined by white dashed contours) run in parallel during adaptation. \textit{Process~1} (the down-sampling and object detection pipeline) operates on sampled configurations, while \textit{Process~2} (the ground-truth pipeline) processes the golden-stream frames required to generate the proxy ground truth during a brief interval within the adaptation period.
\begin{figure*}[t]
\centering
\includegraphics[width=1\textwidth]{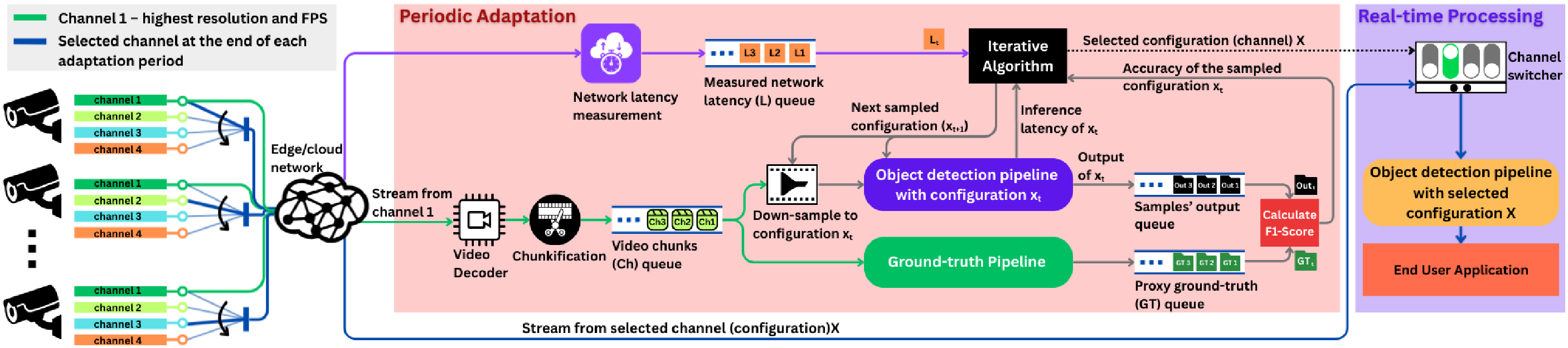}
\caption{Overview of ASTRA, showing how the design in Fig.~\ref{fig:architecture-comparison}(b) is converted into an asynchronous Fork–Join–style pipeline that reduces idle time and improves throughput.}
\label{fig:astra-archit}
\vspace{-10pt}
\end{figure*}

Because Process~1 includes additional steps such as down-sampling and switching between configurations and object detection models, its runtime can fluctuate depending on the chosen configuration. Process~2 runtime can also fluctuate due to variations in available GPU capacity. Consequently, either process can become the bottleneck at different iterations. To prevent one process from stalling the other and to maintain a high throughput, the architecture should execute these concurrent tasks asynchronously by introducing first-in-first-out (FIFO) queues at their inputs and outputs. This transforms the system into a classic \textit{Fork-Join} model, where concurrent processes are decoupled through queued data flow. By structuring the adaptation processes as a \textit{Fork-Join} system with input/output queues, the architecture decouples the concurrent tasks and eliminates global synchronization barriers, which has been shown to significantly improve throughput and resource utilization in parallel processing systems~\cite{marin2020speed,pinto2017understanding}.

In practice, such an asynchronous design can be implemented efficiently using programming constructs such as \textit{futures} and \textit{promises}~\cite{bachan2019upc++,paul2023fine}. These mechanisms enable non-blocking execution of concurrent processes: a \textit{future} represents the result of a computation that is not yet complete, such as a proxy ground-truth, sampled configuration, or measured accuracy, while a \textit{promise} provides a way to deliver that result immediately once it becomes available. Leveraging these constructs allows both pipelines to progress at their own pace while ensuring that the adaptation loop receives synchronized inputs as soon as they are ready, maximizing responsiveness and overall system efficiency. ASTRA is an asynchronous version of the architecture in Fig.~\ref{fig:architecture-comparison}(b) that incorporates this queuing method. This architecture is illustrated in Fig.~\ref{fig:astra-archit}. It incorporates four queues: (i) video chunks queue, (ii) samples' output queue, (iii) proxy ground-truth queue, and (iv) measured network latency queue. 

\noindent\textbf{Chunkification.} To construct the video chunk queue, ASTRA includes a \emph{chunkification} component that receives decoded frames and groups a small number of consecutive frames (e.g., 5) into a single video chunk. A natural choice for the chunk length is the \textbf{I-frame interval}, the number of frames between two successive I-frames in an H264/AVC bitstream. Since all frames within an I-frame interval are decoded together and become available at nearly the same time, aligning the chunk size with the I-frame interval minimizes buffering and idle time. Short I-frame intervals (e.g., 5-6 frames) align naturally with the desired chunk size. Larger intervals, however, require the system to buffer the full interval before generating chunks. For instance, with a 15-frame interval and a 5-frame chunk size, the system must wait for all 15 frames to arrive before producing three chunks. ASTRA trades this minor latency in favor of small, consistent chunk sizes, which are crucial for iterative adaptation algorithm's convergence.

\noindent\textbf{Golden Stream Bandwidth Consumption.} In ASTRA (Fig.~\ref{fig:astra-archit}), the golden stream is transmitted only briefly during each adaptation period. This duration is substantially shorter than the adaptation period, which itself occupies only $\sim$10\% of the time window. In our evaluation setup (Section~\ref{sec:evaluation}), for each camera, the golden stream is transmitted for only $\sim$1\thinspace{s}, while adaptation takes 4--7\thinspace{s} within a 60\thinspace{s} time window. The golden stream consumes approximately 10\thinspace{Mbps} of bandwidth while active. Moreover, ASTRA's modular design allows the adaptation components, particularly \textit{Process~1} (the down-sampling and object detection pipeline) and \textit{Process~2} (the ground-truth pipeline), to be deployed near the cameras with high-bandwidth connectivity, while the continuously running analytics components (the purple box labeled ``Real-time Processing'' in Fig.~\ref{fig:astra-archit}) can connect over bandwidth-constrained links.

\subsection{Latency Measurement}\label{sec:net-lat-meas}
The iterative algorithm requires the total latency (network + inference) for each iteration $t$. While inference latency is measured directly within the object detection pipeline (see Fig.~\ref{fig:astra-archit}), measuring the network latency of video chunk Ch$_t$ is challenging because commodity cameras lack programmable compute units. This measurement is complicated by three factors:

\noindent\textbf{(1) Lack of precise synchronization:} Commodity cameras rely on NTP~\cite{mills1991internet}, which often exhibits clock offsets of \textbf{\emph{tens of milliseconds}} due to network asymmetry~\cite{velez2024assessment,liu2022research}. This margin of error is too large for precise per-chunk latency measurement.

\noindent\textbf{(2) RTP limitations:} The Real-time Transport Protocol (RTP) protocol~\cite{schulzrinne2003rfc3550} is designed for playback jitter control and does not provide a mechanism for the receiver (edge server) to derive one-way network latency without tight clock synchronization.

\noindent\textbf{(3) Variable encoded chunk sizes:} Dynamic compression rates cause encoded chunk sizes to fluctuate unpredictably based on video content and camera model. Consequently, the edge server cannot accurately estimate transmission time based solely on the selected configuration.

\vspace*{2pt}
\noindent\textbf{ASTRA’s estimation of chunk-level network latency.} It is important for the architecture to avoid incorporating latency measurements with high error margins into the adaptation algorithm. Otherwise, the iterative adaptation algorithm may converge to a configuration far from optimal and severely degrade performance. To estimate the end-to-end network latency of each video chunk, ASTRA relies only on receiver-side timestamps at the edge server. Let $t_0$ and $t_1$ denote the times the camera transmits the first and last bits of a video chunk, and $t_2$ and $t_3$
the times the edge receives those bits. Let $d_{\text{first}}$ and $d_{\text{last}}$ denote the one-way path delays experienced by the first and last bits, respectively. By definition $t_2 = t_0 + d_{\text{first}}$ and $t_3 = t_1 + d_{\text{last}}$. The true chunk-level \emph{network latency} that we aim to measure is
\begin{equation}
L_\text{total}^{\text{net}} = t_3 - t_0 = (t_1 - t_0) + d_{\text{last}},
\label{eq:true-net-lat}
\end{equation}
which is equal to the chunk's transmission time plus the one-way network delay of its last bit.

The edge \textbf{\emph{cannot}} observe $t_0$, but it \textbf{\emph{can}} observe $t_3$ and $t_2$, and hence it can measure
\begin{equation}
t_3 - t_2 = (t_1 - t_0) + (d_{\text{last}} - d_{\text{first}}),
\label{eq:observable}
\end{equation}
Since the chunk sizes are small (\(\sim\!0.2\)\thinspace{s}) the jitter term \((d_{\text{last}} - d_{\text{first}})\) is negligible, making
\begin{equation}
t_3 - t_2 \approx t_1 - t_0.
\label{eq:approx_time}
\end{equation}
Thus, using \eqref{eq:true-net-lat} and \eqref{eq:approx_time} and the fact that one-way delay $d_{\text{last}}$ is roughly half of Round Trip Time (RTT), ASTRA approximates the per-chunk network latency as:

\begin{equation}
L_\text{total}^{\text{net}} \approx (t_3 - t_2) + \frac{\text{RTT}}{2},
\label{eq:astra-estimate}
\end{equation}

The error introduced by assuming a small jitter term \(d_{\text{last}} - d_{\text{first}}\) and by approximating one-way delay as RTT/2 is much smaller than the tens of milliseconds errors introduced by NTP timestamps~\cite{velez2024assessment,liu2022research} or RTP sender timestamps. This way, ASTRA obtains a practical and sufficiently accurate per-chunk network-latency estimate using only receiver-side timestamps without requiring any programmable device on the camera. To measure RTTs, ASTRA's \emph{network latency measurement} component sends periodic ping requests to the camera during the transmission of each video chunk and computes the average RTT. 

\section{Configuration Switching in Object Detection Pipeline}
In addition to the overhead of switching video stream configurations, the frequent switching of configurations in the object detection pipeline (dark purple box in Fig.~\ref{fig:astra-archit}) also incurs non-negligible overhead. As discussed in Sec.~\ref{sec:online-adaptation} and Sec.~\ref{subsec:astra-architect}, the adaptation iterations needs to occur back-to-back in short time slots (e.g., \(\sim\!0.2\)\thinspace{s}) which requires high frequency of switching between configurations in the object detection pipeline (dark purple box in Fig.~\ref{fig:astra-archit}). In this section, we discuss the overheads of configuration switching in this pipeline and how these overheads can be mitigated.

To understand inherent delays of configurations switching in the object detection pipeline, we conducted a thorough profiling of the pipeline at runtime. The results of our profiling, combined with the measurements presented in previous studies such as~\cite{yi2020analysis,yan2022dissecting,yao2022eais}, reveal that the primary overhead of such switching includes: 

\noindent\textbf{Pipeline reconfiguration overheads}: Processing video chunks with different resolutions and frame rates requires recompiling and initializing pipeline elements. For example, switching to a configuration with a different model variant or input shape requires reloading and initializing the inference engine and allocating model-specific buffers which can take \textbf{\emph{tens to hundreds of milliseconds}} per switch.

\noindent\textbf{Memory allocation and deallocation overheads}: Processing a sequence of video chunks with different configurations triggers frequent memory allocation and deallocation for intermediate buffers across the pre-processing, inference, and post-processing stages, introducing additional delays of \textbf{\emph{a few to tens of milliseconds}}.

\noindent\textbf{Cache inefficiencies}: Switching between video chunks and configurations can disrupt cache continuity, resulting in increased cache misses. Since GPU performance is highly dependent on the memory hierarchy, these cache misses lead to less efficient memory access and extra delays~\cite{candel2018improving,sepanski2022maximizing} (\textbf{\emph{a few to tens of milliseconds}}).

These delays, even though small, mostly ranging from a few to hundreds of milliseconds, could severely impact the adaptation when happening recurrent. Consider an adaptation algorithm that can determine a near-optimal configuration in ten iterations. With a total switching delay of even a few hundred milliseconds per iteration, the system would spend several seconds solely on configuration switching in the object detection pipeline.

\subsection{Mitigating Switching Overhead}\label{sec:mitigate-switching-delay}
We applied the following optimizations to the dark purple \emph{Object detection pipeline} in Fig.~\ref{fig:astra-archit}, reducing switching latency from hundreds of milliseconds to 10-20\thinspace{ms}. 

\noindent\textbf{Efficient buffer management}: Since video chunks are small, large inter-element queues only add switching delay. We cap these buffers to one frame and reuse small buffer pools to reduce draining, flushing, and reallocation overhead.

\noindent\textbf{Device memory versus host memory}: In ASTRA's object detection pipelines (dark purple and green boxes in Fig.~\ref{fig:astra-archit}), frames remain in GPU device memory throughout the hot path (decoder $\rightarrow$ preprocess $\rightarrow$ DNN $\rightarrow$ postprocess). We use NVMM/device-resident buffers and custom kernels on \texttt{cudaMalloc} pointers, avoiding CUDA zero-copy host memory because PCIe reads add higher per-chunk latency\footnote{We have published measurements of memory-type impact on latency in~\cite{ghasemi2025pave}}.

\noindent\textbf{Dynamic pipeline reconfiguration}: ASTRA reconfigures the object-detection pipeline at runtime without recreating elements, reducing switching delay from hundreds of milliseconds to about ten milliseconds. It uses dynamic CPU/GPU memory allocation to update configuration-dependent state and preloads all supported TensorRT engines into GPU memory so switching only selects an already initialized model. Idle models use memory but add no computation.

\section{Evaluations}\label{sec:evaluation}
In this section, we evaluate ASTRA’s architecture in terms of accuracy, latency, and GPU utilization under diverse network conditions and video contents. We plug two iterative algorithms into the architecture (the "content" of the black box in Fig.~\ref{fig:astra-archit}): (i) GP-UCB-C derived from~\cite{xu2023constrained} and (ii) Chameleon++ derived from~\cite{jiang2018chameleon}. These algorithms represent two common classes of adaptation methods: (1) theoretically guaranteed algorithms (GP-UCB-C) and (2) empirically estimated algorithms (Chameleon++). We compare their performance against two baselines: (a) the offline optimal (upper bound on achievable performance) and (b) a ground-truth-free adaptation approach that estimates accuracy using confidence scores rather than ground truth.

As described in Section~\ref{sec:online-adaptation}, the goal of the \emph{iterative algorithm} is to select a configuration $x \in \mathcal{X}$ that maximizes a weighted accuracy--latency utility,
\[
\max_{x \in \mathcal{X}} \; c_a A(x) - c_l L(x)
\quad \text{s.t.} \quad 
A(x) \ge \alpha,\;\; L(x) \le \beta,
\tag{P1}
\label{eq:setup}
\]
where $A(x)$ and $L(x)$ denote average accuracy and latency, $c_a, c_l \ge 0$ are application-specific weights, and $\alpha,\beta$ are application-specific accuracy and latency constraints.

\noindent\textbf{GP-UCB-C:}
Following the constrained Bayesian optimization method introduced in~\cite{xu2023constrained}, we model the performance of each configuration using Gaussian Processes (GPs)~\cite{srinivas2009gaussian}. Specifically, the accuracy function $A(x)$ and latency function $L(x)$ over the discrete configuration set $\mathcal{X}$ are treated as correlated, unknown functions. For each configuration $x$ and iteration $t$, we denote by $\mathrm{UCB}^A_t(x)$ and $\mathrm{LCB}^A_t(x)$ (resp. $\mathrm{UCB}^L_t(x)$ and $\mathrm{LCB}^L_t(x)$) the upper and lower confidence bounds on the accuracy $A(x)$ (resp. latency $L(x)$) given by the GP posterior. After each observation, the GP posterior provides predictive means $\mu_t^A(x)$, $\mu_t^L(x)$ and confidence intervals $[\mathrm{LCB}_t^A(x), \mathrm{UCB}_t^A(x)]$ and $[\mathrm{LCB}_t^L(x), \mathrm{UCB}_t^L(x)]$ for accuracy and latency, respectively.

At each iteration $t$, the algorithm selects the next configuration $x_t$ by maximizing the UCB on the utility,
\[
x_t = \arg\max_{x \in \mathcal{A}_t}
\; c_a\, \mathrm{UCB}^A_t(x) - c_l\, \mathrm{LCB}^L_t(x),
\]
over the feasible set
\[
\mathcal{A}_t = \left\{x \in \mathcal{X}
\;\middle|\;
\mathrm{UCB}^A_t(x) \ge \alpha,\;
\mathrm{LCB}_t^L(x) \le \beta
\right\}.
\]
Configurations that are likely to violate the accuracy or latency constraints are therefore removed from $\mathcal{A}_t$.

After evaluating $A_t(x_t)$ and $L_t(x_t)$, the GP posterior is updated and the feasible set $\mathcal{A}_{t+1}$ is recomputed. The algorithm then identifies the current best feasible configuration
\[
h_t = \arg\max_{x \in \mathcal{A}_t} \; c_a \mu_t^A(x) - c_l \mu_t^L(x)
\]
and the most promising alternative
\[
r_t = \arg\max_{x \in \mathcal{A}_t \setminus \{h_t\}} \; c_a\, \mathrm{UCB}^A_t(x) - c_l\, \mathrm{LCB}^L_t(x).
\]
The procedure stops at iteration $t$ when the lower confidence bound on the utility of $h_t$ is no smaller than the upper confidence bound on the utility of $r_t$, i.e.,
\[
c_a\, \mathrm{LCB}^A_t(h_t) - c_l\, \mathrm{UCB}^L_t(h_t)
\;\ge\;
c_a\, \mathrm{UCB}^A_t(r_t) - c_l\, \mathrm{LCB}^L_t(r_t),
\]
and returns $h_t$ as the selected configuration.

We refer to this constrained UCB-based selection rule, together with the feasibility update and the early stopping condition above, as \textbf{GP-UCB-C}.

\vspace*{2pt}
\noindent\textbf{Chameleon++:} 
This algorithm is an improvement of the Chameleon algorithm~\cite{jiang2018chameleon} that adds latency-awareness while preserving the original search logic. Chameleon is a profiling-based configuration search algorithm. Namely, it measures the accuracy of a small subset of configurations. It assumes that configuration parameters (e.g., resolution and frame rate) contribute independently to accuracy in order to estimate the accuracy of all remaining configurations, without exhaustive profiling. Chameleon++ extends this approach by also measuring inference latency during profiling. For configurations that are not profiled, their latency is estimated using the same parameter-independence assumption used by Chameleon for accuracy estimation. This allows Chameleon++ to jointly consider accuracy and latency without modifying Chameleon's core algorithm. 

\noindent\textbf{AutoML++ (ground-truth-free adaptation):} In this benchmark, the iterative algorithm is the GP-UCB-C algorithm that uses confidence scores generated by the object detection model as the accuracy metric instead of F1-scores that need proxy ground truth. This approach is inspired by the AutoML adaptation platform~\cite{galanopoulos2021automl}, a GP-UCB-based method that uses confidence score as the accuracy metric.

\noindent\textbf{Offline optimal:} This benchmark uses, for each time window, the optimal configuration given full offline knowledge of $A(x)$ and $L(x)$. For multi-camera experiments, $A(x)$ and $L(x)$ in (P1) are computed as averages over all cameras. The optimal configuration is found by exhaustively evaluating all configurations over the entire time window and selecting the best feasible one according to (P1). Importantly, optimality is defined with respect to (P1), including its accuracy and latency constraints, rather than with respect to either metric individually. For example, in the accuracy-constrained setting, the offline optimal is the lowest-latency configuration that satisfies the minimum-accuracy threshold. An online method may show lower latency only by violating the corresponding accuracy threshold. Therefore, appearing better on one individual metric does not indicate a better feasible solution to P1. This benchmark has no adaptation period, incurs no adaptation overhead, and serves as an upper bound benchmark on achievable performance.

\vspace*{2pt}  
\noindent\textbf{Object detection models.}
In our experiments, the adaptation is performed for the object detection task, specifically pedestrians and vehicles detection, using YOLOv8~\cite{terven2023comprehensive} (Section~\ref{sec:cosmos-deployment}) and YOLOv4~\cite{bochkovskiy2020yolov4} (Sections~\ref{sec:google-cloud-eval} and~\ref{subsec:eva-varing-net}) models. Importantly, ASTRA is model-agnostic: any DNN-based object detection model, including newer versions of YOLO, can be integrated into its architecture. 

\vspace*{2pt}
\noindent\textbf{Optimization setup and parameters.} The evaluations are performed for two sets of parameters in (P1): (i) \textit{\textbf{accuracy constrained}}, i.e., $c_a = 0$, $c_l = 1$, $\beta = +\infty$, and $\alpha \gg 0$ (Sections~\ref{sec:cosmos-deployment} and~\ref{sec:google-cloud-eval}) and (ii) \textit{\textbf{latency constrained}}, i.e., $c_a = 1$, $c_l = 0$, $\alpha = 0$, and $\beta \ll +\infty$ (Section~\ref{subsec:eva-varing-net}). The configuration set $\mathcal{X}$ used consists of $12$ elements, each of which is a tuple $(r_i,f_i)$ where $r_i$ is one of the values $\{832,608,416\}$ and $f_i$ is one of the values $\{30,15,10,5\}$. A video stream with configuration $(r,f)$ has resolution $r\times r$ and frame rate $f$ fps. These three resolution levels and four frame-rate levels provide sufficiently distinct operating points for the adaptation algorithm. Configurations with only small differences in resolution or frame rate typically exhibit similar accuracy and latency, and therefore using a much denser configuration space would increase the adaptation search space with limited additional performance diversity. Increasing the number of sufficiently distinct candidate configurations can increase adaptation time by enlarging the search space.

All experiments use an F1-score threshold of \textbf{0.7} and chunk size $\tau$=\textbf{0.2\thinspace{s}} (also referred to as time slot). The time window size is determined by the time required for adaptation and the video content/network volatility. A smaller time window is needed for higher volatility, but it is essential to keep time window size notably larger than the adaptation period, e.g., $\sim$10x (see Fig.~\ref{fig:ideal-adapt-timeline}). ASTRA's adaptation period is typically \textbf{4-7\thinspace{s}} when considering \textbf{12} configurations. Accordingly, \emph{\textbf{time window length of 60\thinspace{s}}} is used in all the experiments. Consequently, for a 60-second time window and 0.2-second slot duration, the total number of time slots is $T=\frac{60}{0.2}=300$. 

\vspace*{2pt}
\noindent\textbf{Accuracy and latency per time window.} During the adaptation period of camera $i$, in the $n$-th time window, the system (object detection pipeline denoted by light orange box in Fig.~\ref{fig:astra-archit}) uses the configuration chosen in the $(n-1)$ time window (denoted by $x^{(i,n-1)}$). For the remainder of the time window, it uses the configuration chosen by the adaptation algorithm for time window $n$ (i.e., $x^{(i,n)}$). Therefore, for $K$ cameras, the average achieved accuracy and latency at time window $n$ are calculated as follows:
\begin{equation}\label{eq:acc-lag-mutli-cam-def}
    \arraycolsep=1.4pt\def\arraystretch{1.4}
    \begin{array}{rl}
        A_n=& \cfrac{1}{K} \sum\limits_{i=1}^K \cfrac{1}{T} \left(\sum\limits_{t=1}^{\eta_{i,n}} A_{t,i}(x^{(i,n-1)}) + \sum\limits_{t=\eta_{i,n} + 1}^{T} A_{t,i}(x^{(i,n)})\right)   \\[2ex]
        L_n=& \cfrac{1}{K} \sum\limits_{i=1}^K\cfrac{1}{T} \left(\sum\limits_{t=1}^{\eta_{i,n}} L_{t,i}(x^{(i,n-1)}) + \sum\limits_{t=\eta_{i,n}+1}^T L_{t,i}(x^{(i,n)})\right) 
    \end{array} 
\end{equation}
where $T$ is the total number of slots in a time window, and $\eta_{i,n}$ is the number of slots in the adaptation period of camera $i$ within the $n$\textsuperscript{th} window. For camera $i$, $A_{t,i}(x^{(i,n-1)})$ and $L_{t,i}(x^{(i,n-1)})$ denote the accuracy and latency achieved using configuration $x^{(i,n-1)}$ at time slot $t$.


\vspace*{2pt}
We performed evaluations for two settings: (i) real-time adaptation of two live cameras within the COSMOS testbed (Section~\ref{sec:cosmos-deployment}), and (ii) real-time adaptation of up to eight emulated cameras using a comprehensive video dataset on the Google Cloud Platform (Sections~\ref{sec:google-cloud-eval} and~\ref{subsec:eva-varing-net}).

\begin{figure}[t]
    \centering
    \includegraphics[width=1\linewidth]{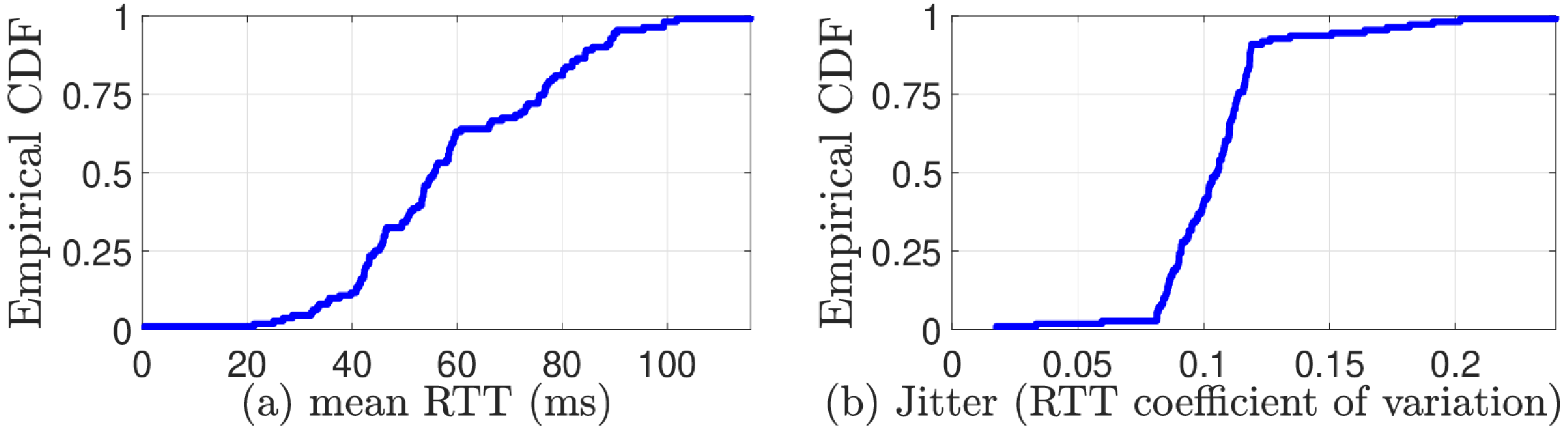}
    \caption{CDF of (a) mean RTT and (b) jitter from end users to the nearest Alibaba Cloud sites under 4G~\cite{xu2021cloud}.} 
    \label{fig:xu-dataset}
\vspace{-10pt}    
\end{figure}

\vspace*{2pt}
\noindent\textbf{Network latency dataset.} In both settings, the live and emulated cameras are connected to the servers via high-speed links (e.g., COSMOS cameras use fiber optics). This ideal connectivity is atypical for distributed edge computing systems. Therefore, to evaluate the performance of ASTRA under realistic network conditions between the cameras and the edge servers, we used a dataset of extensive real network latency measurements provided in~\cite{xu2021cloud} to induce extra latency. The dataset contains the RTT between end users and the nearest Alibaba Cloud site under 4G. 
The empirical Cumulative Distribution Function (CDF) of the mean RTT and jitter (RTT coefficient of variation measured as the standard deviation divided by the empirical mean) are presented in Fig.~\ref{fig:xu-dataset}. The detailed dataset of these measurements was graciously provided by the authors of~\cite{xu2021cloud}.
\begin{figure*}[t]
\centering
\includegraphics[width=0.4\textwidth]{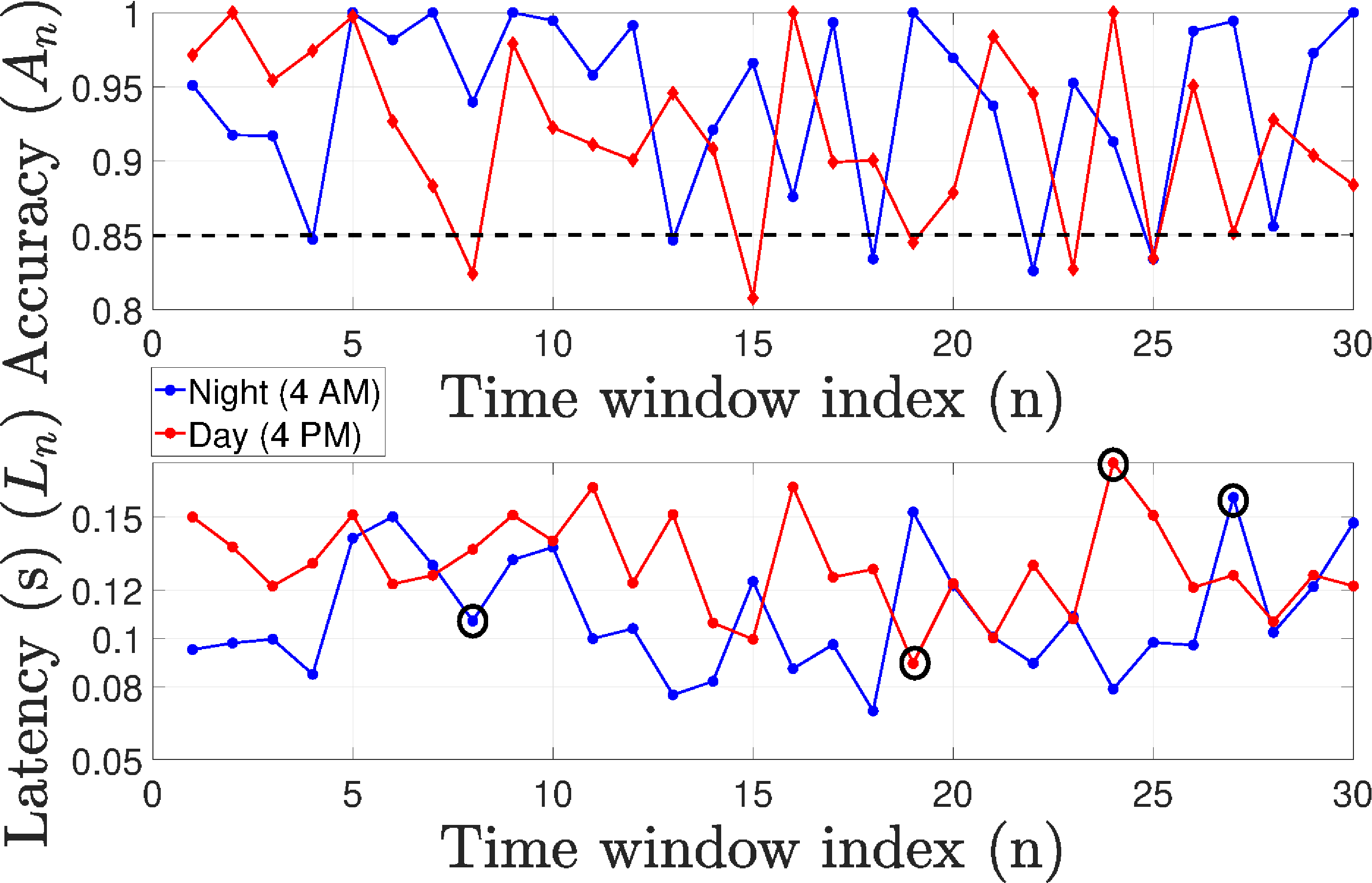}
\hspace{0.2 in}
\includegraphics[width=0.42\textwidth]{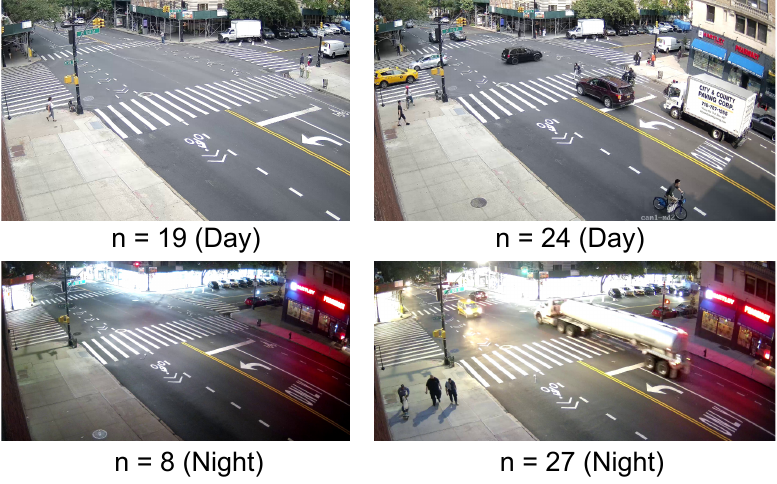}
\caption{Left: ASTRA's (with GP-UCB-C as the algorithm) achieved average accuracy and latency at each time window, during two 30-minute periods (peak/day and off-peak/night). Right: Snapshots of the intersection at circled time windows. 
}
\label{fig:cosmos-trace}
\vspace{-10pt}
\end{figure*} 
\subsection{Deployment in COSMOS Testbed}\label{sec:cosmos-deployment}
We deployed ASTRA within the COSMOS testbed~\cite{yang2020cosmos} and assessed its adaptation capabilities using two \emph{live} cameras. This deployment validates the system's real-time performance under actual physical constraints. This effectively addresses a gap in prior research, where evaluations involving live cameras are rarely conducted. COSMOS cameras are located on the 1$\textsuperscript{st}$ and 2$\textsuperscript{nd}$ floor of a building, viewing 120$^\textrm{th}$ St.\ and Amsterdam Ave.\ intersection, NYC (see Fig.~\ref{fig:cameras-view}). Moreover, two COSMOS servers running Ubuntu 20.04 with Intel Xeon CPU@2.60GHz and an NVIDIA V100 GPU, were used to deploy ASTRA. One server hosts the \emph{real-time processing} (denoted by the light purple container in Fig.~\ref{fig:astra-archit}) and the other hosts the \emph{periodic adaptation} (denoted by the pale red container in Fig.~\ref{fig:astra-archit}). We emulated realistic network latencies between cameras and the servers using the dataset described above.

\noindent\textbf{Parameter tuning ($\alpha$).} We experimented with $\alpha$ ranging from 0.4 to 1 (ideal). We observed that values below 0.85 compromised pedestrian detection accuracy, while values above 0.85 incurred high latency and GPU utilization. Consequently, we selected $\alpha=0.85$ as the threshold for the \textit{\textbf{accuracy-constrained}} scenario to balance accuracy and overhead. 

We executed ASTRA with the GP-UCB-C algorithm during two 30-minute periods, representing peak (4 PM) and off-peak (4 AM) hours. Fig.~\ref{fig:cosmos-trace} shows the average achieved accuracy, $A_n$, and latency, $L_n$, for each time window, as defined in~\eqref{eq:acc-lag-mutli-cam-def}. The results confirm that ASTRA consistently maintains accuracy above or close to the required threshold of 0.85 in both periods. Notably, latency is lower during off-peak hours, as reduced pedestrian and vehicle density allows the system to select lower resolutions and frame rates. To illustrate ASTRA's adaptability, we provide snapshots of the intersection at specific timestamps. For instance, at window $n=19$ (day), ASTRA leveraged the low traffic density to reduce the resolution/frame rate, lowering latency while meeting the accuracy target. Conversely, at $n=24$, increased density and speed prompted ASTRA to upgrade the configuration to preserve accuracy, resulting in higher latency. A similar adaptive pattern is observed at night ($n=8$ and $n=27$). 
\subsection{Controlled and Reproducible Evaluation}\label{sec:google-cloud-eval} 
To thoroughly evaluate ASTRA and ensure a fair comparison of iterative algorithms, it is necessary to use identical and reproducible conditions, such as consistent video content and network conditions. Therefore, we deployed ASTRA in Google Cloud VM instances and emulated multiple cameras using a comprehensive video dataset. Experiments were conducted using three VM instances hosted in the \texttt{us-east1} region (Moncks Corner, South Carolina). Two of these instances were equipped with an NVIDIA T4 GPU and were used to emulate the \emph{real-time processing} (denoted by the light purple container in Fig.~\ref{fig:astra-archit}) and the \emph{periodic adaptation} (denoted by the pale red container in Fig.~\ref{fig:astra-archit}). The third instance was dedicated to streaming H264-encoded videos over RTP (RTSP over TCP) using GStreamer, thereby emulating live cameras. 

This setup contrasts with prior studies (e.g.,~\cite{du2023oneadapt,wong2024madeye,zhang2024starstream,wang2023shoggoth,zhao2025mocha}), which typically rely on reading pre-recorded video files directly from local storage. By streaming videos over the network, we subject ASTRA to real-time latency and throughput constraints, whereas file-based approaches do not capture these challenges inherent to live video ingestion. To emulate realistic network latencies between the (emulated) cameras and the edge servers, we used the network latency dataset introduced earlier in this section. 

\noindent\textbf{{Video dataset.}} The dataset consists of videos recorded simultaneously from the two COSMOS testbed's cameras (used in Section~\ref{sec:cosmos-deployment}). It includes 52 pairs of ten-minute videos (104 in total). Each pair is recorded from the two cameras simultaneously. The video pairs are captured eight times daily from morning to evening, in June, July, and Sept.\ 2022. The videos were purposely selected to represent a range of traffic densities and weather conditions (sunny, rainy, etc.).

\begin{figure*}[t]\label{fig:multi-camera}
\centering
\includegraphics[width=0.9\textwidth]{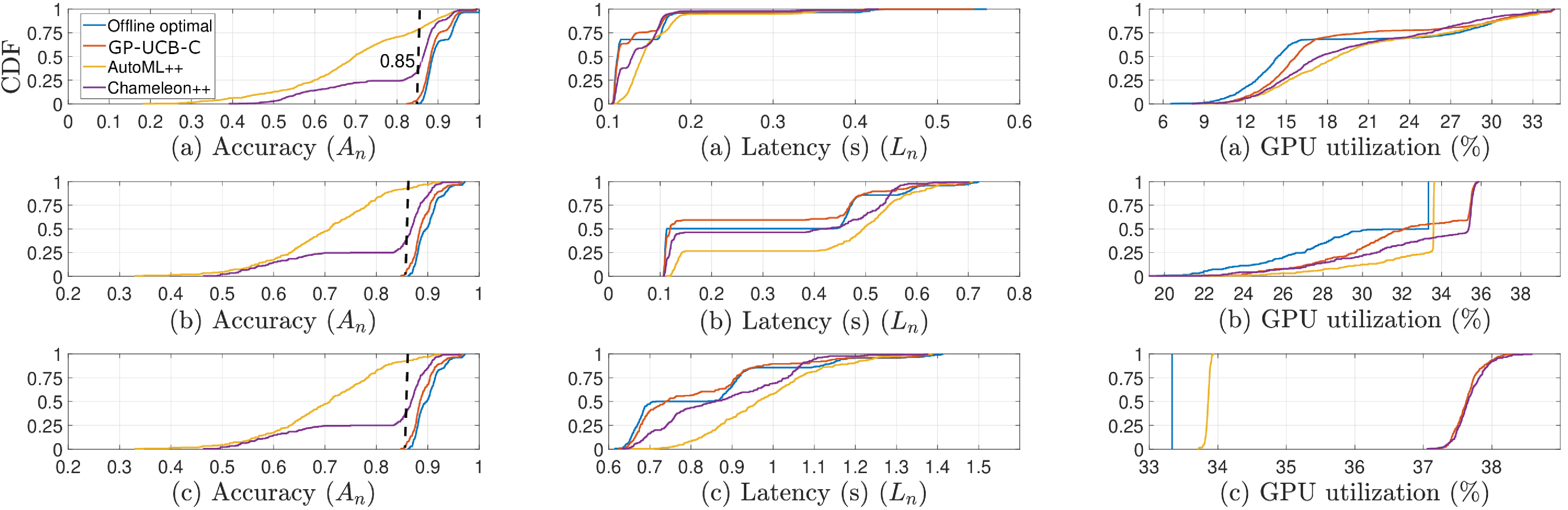}
\caption{Empirical CDF of average achieved accuracy (with lower bound of 0.85), average achieved latency, and average GPU utilization per time window for the setting with (a) two, (b) four, and (c) eight cameras.}
\label{fig:4-cam}
\end{figure*} 
\begin{figure*}[t]
\centering
\includegraphics[width=1\linewidth]{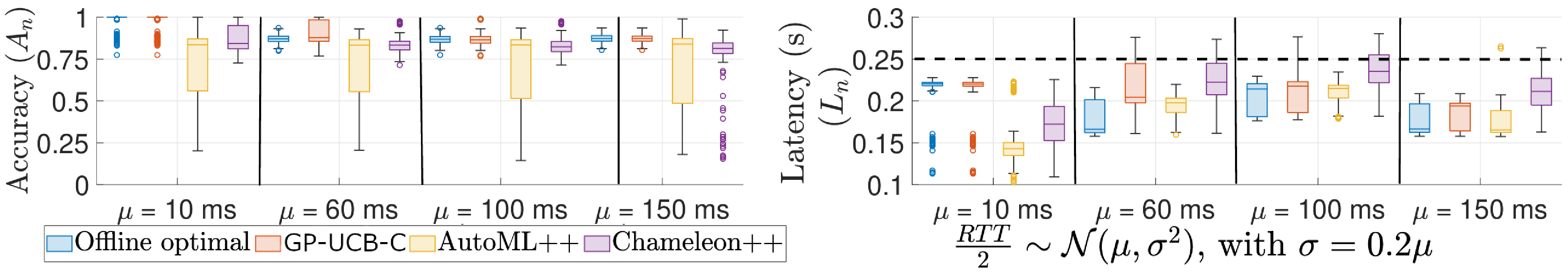}
\caption{Performance comparison under varying network latency conditions.}
\vspace{-10pt}
\label{fig:box-plot}
\end{figure*} 
\noindent\textbf{Multi-camera.}
In our setting, in a single instance of ASTRA, real-time adaptation of one camera takes 4--7\thinspace{s}. Increasing the number of cameras does not increase the per-camera configuration search space, but increases the aggregate adaptation workload of an ASTRA instance, since each camera requires an independent adaptation process. Depending on the available adaptation resources, cameras can be adapted individually or in batches. In our evaluation, cameras are adapted sequentially, with each camera adapted once during a repeating 60\thinspace{s} time window and with a different adaptation offset within the window. Therefore, a 60\thinspace{s} time window allows for $\lfloor60/7\rfloor=8$ cameras. To increase the number of cameras, multiple instances of ASTRA architecture with similar performance can be set up, thereby scaling/multiplying the 8-camera system. Accordingly, we conducted adaptation for up to eight emulated cameras using all the benchmark algorithms for the accuracy-constrained scenario. Each emulated camera is associated with a set of videos (26 videos for four or more emulated cameras and 52 videos for two emulated cameras). \textbf{\emph{GPU utilization}} was measured as the average utilization over both GPU-equipped edge servers in the entire time window.

\vspace*{1pt}
\noindent\textbf{Performance comparison.} Fig.~\ref{fig:4-cam} shows the system-level impact of increasing the number of cameras, and consequently the aggregate adaptation workload, from 2 to 4 and 8 cameras. Despite this increase, \textit{compared to offline optimal}, ASTRA with GP-UCB-C algorithm achieves near-optimal performance (less than 10\% deviation) with GPU utilization around 2\%, 4\%, and 9\% for 2, 4, and 8 cameras, respectively, which means around 2\% per camera. Moreover, it satisfies the accuracy threshold of 0.85 in more than 90\% of the time in all cases. Note that an online method can occasionally appear better than the offline optimal on one individual metric ((in this case, lower latency), but such cases correspond to a violation of the associated constraint (in this case, minimum accuracy of 85\%) and therefore do not represent a better feasible solution to P1.

GPU utilization of ASTRA with Chameleon++ is close to GP-UCB-C. However, Chameleon++ does not meet the specified accuracy target, $\alpha=$0.85, in more than 30\% of the times across all the evaluated scenarios. AutoML++ (GPU-UCB-C with no ground-truth) has the lowest GPU utilization as it does not require proxy ground truth. Despite having low GPU overhead, AutoML++ fails to achieve the minimum accuracy required around 75\%, 85\%, and 90\% of times for the 2, 4, and 8 camera scenarios, respectively. This confirms the importance of incorporating proxy ground-truth for adaptation.

\subsection{Varying Network Latencies}\label{subsec:eva-varing-net}
To assess adaptation robustness to varying network conditions, we conducted another experiment with increased latency and jitter. The measured latencies in~\cite{xu2021cloud} and similar network latency measurement studies~\cite{dang2021cloudy,schlinker2019internet} are relatively small. Therefore, following the method used in~\cite{zhang2018awstream} to evaluate AWStream, we added Gaussian-distributed delay with a mean between 10\thinspace{ms} and 150\thinspace{ms} and a variance of 20\% between the emulated camera and the edge servers. To determine how effectively the GP-UCB-C algorithm and other benchmarks adapt to network variation, we consider \textit{\textbf{latency constrained}} optimization with $\beta$ = 250\thinspace{ms}.

The results of these experiments are presented in Fig.~\ref{fig:box-plot}. As can be observed, the performance of AutoML++ is unstable under varying network conditions. In contrast, the performance of Chameleon++ and GP-UCB-C is more consistent. Notably, GP-UCB-C's performance is closest to the optimal and demonstrates the least variation.

These evaluations show that ASTRA, when instantiated with a principled iterative algorithm such as GP-UCB-C, can track the best configuration under dynamic workloads with minimal overhead. Across single- and multi-camera experiments, ASTRA consistently approaches the performance of an offline optimal benchmark that has full knowledge of accuracy and latency and incurs no adaptation overhead, demonstrating that fast online adaptation is both feasible and effective.

\noindent\textbf{Relationship to Network Conditions.}
The experiments in Fig.~\ref{fig:box-plot} explicitly vary network latency and jitter. However, reduced bandwidth and packet loss can affect ASTRA in a similar way by increasing the time required for a video chunk to arrive at the edge server. Since ASTRA receives video streams over TCP, reduced available bandwidth increases chunk transmission and queueing time, while packet loss can trigger retransmissions and further delay chunk delivery. Throughput limitations may also reduce the delivered video quality, which can lower the measured analytics accuracy. Therefore, the effects of adverse network conditions are reflected in ASTRA's measured chunk-level latency and accuracy and, consequently, in the configuration selected by the adaptation algorithm. Thus, the experiments in Fig.~\ref{fig:box-plot} evaluate ASTRA's response to increased chunk-delivery latency, which is a common consequence of several adverse network conditions, including increased propagation delay, jitter, bandwidth limitations, and packet loss.


\section{Conclusions}\label{sec:conclusion}
We presented ASTRA, a lightweight architecture for executing a broad class of iterative adaptation algorithms for real-time video analytics. ASTRA enables fast configuration adaptation with minimal overhead and supports algorithms ranging from theoretically grounded methods to empirical, profile-driven approaches. Our evaluations show that ASTRA, when paired with a principled algorithm, consistently tracks near-optimal configurations and achieves performance close to an offline optimal benchmark that has full knowledge of accuracy and latency and incurs no overhead (upper bound on achievable performance). These results demonstrate that principled online adaptation is both practical and effective for live video analytics systems. \textbf{Future work} will focus on several directions: (i) scaling live evaluations to significantly larger camera networks, (ii) developing adaptive time windows to better handle rapid environmental fluctuations, and (iii) extending ASTRA to large-scale geo-distributed edge/cloud systems with dynamic resource allocation.  

\section*{Acknowledgment}
This work was supported in part by NSF Center for Smart Streetscapes (CS3) under NSF Cooperative Agreement EEC-2133516, NSF grant CNS-2038984 and corresponding support from the Federal Highway Administration (FHWA), NSF grant CNS-2148128 and by funds from federal agency and industry partners as specified in the Resilient \& Intelligent NextG Systems (RINGS) program, NSF grant CNS-2450567, and ARO grants W911NF2210031 and W911NF1910379.


\bibliography{ref.bib}

\end{document}